\documentclass[reprint,
					aps,
					pra,
					english,
					nofootinbib,
					10pt
]{revtex4-1}

	\usepackage[utf8]{inputenc}  

	\usepackage[T1]{fontenc}  

	\usepackage[intlimits]{amsmath}		

	\usepackage{amssymb}

	\usepackage{mathtools}		

	\usepackage{dsfont}      

	\usepackage{fixmath}     

	\usepackage[caption=false]{subfig}

	\usepackage{graphicx}

	\usepackage{ifthen}

	\usepackage[final,colorlinks=true,linkcolor=blue,citecolor=blue]{hyperref}    

  \renewcommand{\d}[1]{\mathrm{d}#1}
  \newcommand*\dtwo[1]{\mathrm{d}^2#1}
  \newcommand{\ii}{\mathrm{i}}
  \newcommand*{\unity}{\mathds{1}}
  \newcommand{\conj}[1]{\xoverline{#1}}
  \newcommand{\abs}[1]{\left\lvert{#1}\right\rvert}
  \newcommand{\ee}[1]{\operatorname{e}^{#1}}
  \newcommand*\defeq{\equiv}
  
  \newlength{\bracewidth}
  \newcommand{\myunderbrace}[2]{\settowidth{\bracewidth}{$#1$}#1\hspace*{-1\bracewidth}\smash{\underbrace{\makebox{\phantom{$#1$}}}_{#2}}}
  \newcommand*{\per}{\operatorname{per}}

  \newcommand{\bra}[1]{\langle #1 \vert}
  \newcommand{\ket}[1]{\vert #1 \rangle}
  \newcommand*\ketbra[2]{\ket{#1}\bra{#2}}

\renewcommand{\conj}[1]{#1^*}
\renewcommand*\d[1]{d #1\,}

	\newcommand*\Mports{M}
	\newcommand*\Ndets{N}
	\newcommand*\permut{\sigma}
	\newcommand*\permutd{\delta}
	\newcommand*\PathGroup{\Omega^{(\sourceSet)}}
	\newcommand*\detd{d}
	\newcommand*\sources{s}
	
	\newcommand*\Econstant{K}

	\newcommand*\SymmGroup[1]{\Sigma_{#1}}

\newcommand*\density[1]{\hat{\rho}_{#1}}

\newcommand*\fdist{\xi}
\newcommand*\tdist{\chi}

\newcommand*\meanrate[1]{\bar{r}_{#1}}
\newcommand*\nmean[1]{\bar{n}_{#1}}
\newcommand*\Pfunct{P}

\newcommand*\sourceSet{\mathcal{S}_{\Ndets}}

\newcommand*\sourcevec[1]{\sources_{#1}}

\newcommand*\Npart[1]{N_{#1}(\sourceSet)}

\newcommand*\ProdMatrix[2]{\mathcal{C}^{(\setD,\sourceSet)}_{#2}}
\newcommand*\thermaloverlap[3]{\tdist_{#1}(\tout{#3} - \tout{#2})}

\newcommand*{\ain}[1]{\hat{a}_{#1}}

	\newcommand*\Eoutchar{\mathrm{E}}
	\newcommand*\Eout[1]{\hat{\Eoutchar}_{#1}(\tout{#1})}
	\newcommand*\Eoutplus[1]{\hat{\Eoutchar}^{(+)}_{#1}(\tout{#1})}
	\newcommand*\Eoutminus[1]{\hat{\Eoutchar}^{(-)}_{#1}(\tout{#1})}
	\newcommand*\Eoutplusminus[1]{\hat{\Eoutchar}^{(\pm)}_{#1}(\tout{#1})}
	
	\newcommand*\Einchar{E}
	\newcommand*\Einplus[2]{\hat{\Einchar}^{(+)}_{#1}\ifthenelse{\equal{\unexpanded{#2}}{}}{}{(\tout{#2})}}
	\newcommand*\Einminus[2]{\hat{\Einchar}^{(-)}_{#1}\ifthenelse{\equal{\unexpanded{#2}}{}}{}{(\tout{#2})}}

\newcommand*{\Umatrix}[1]{\mathcal{U}^{(\setD,#1)}}
\newcommand*{\UmatrixTot}{\mathcal{U}}
\newcommand*{\UmatrixRect}{\mathcal{U}^{(\setD)}}
\newcommand*{\UmatrixRectAdj}{\mathcal{U}^{\dagger(\setD)}}
\newcommand*\Uto[2]{\mathcal{U}_{#2,#1}}
\newcommand*\Utoconj[2]{\conj{\mathcal{U}}_{#2,#1}}

\newcommand*\Amatrix{\mathcal{A}^{(\setD)}}
\newcommand*\AmatrixEl[2]{\mathcal{A}_{#1,#2}}
\newcommand*\Cmatrix{\mathcal{B}^{(\setD)}_{\toutSetShort}}

\newcommand\tout[1]{t_{#1}}
\newcommand*\toutSet{ \left\{ \tout{\detd} \right\}_{\detd\in\setD}}
\newcommand*\toutSetShort{ \left\{ \tout{\detd} \right\}}
\newcommand*\setD{\mathcal{D}_{\Ndets}}
\DeclareMathOperator\tr{tr}

	\newcommand*\Gn[1]{G^{(N)}_{#1}(\toutSetShort; \setD)}
	\newcommand*\Gnbare[1]{G^{(N)}_{#1}}
	\newcommand*\Gone[2]{G^{(1)}_{#1}(\tout{#2},\tout{#2})}
	\newcommand*\Gonebare[1]{G^{(1)}_{#1}}

	\newcommand*\Gonein[3]{\mathcal{G}^{(1)}_{#1}(\tout{#2},\tout{#3})}

\begin{document}
\title{Multi-Boson Correlation Interferometry with Multi-Mode Thermal Sources}
\author{Vincenzo Tamma}
\author{Simon Laibacher}
\affiliation{Institut f\"{u}r Quantenphysik and Center for Integrated Quantum Science and Technology (IQ\textsuperscript{ST}), Universität Ulm, D-89069 Ulm, Germany}

\begin{abstract}
We develop a general description of multi-boson interferometry based
on correlated measurements in arbitrary passive linear interferometers for multi-mode thermal
sources with arbitrary spectral distributions. The multi-order
correlation functions describing the multi-boson detection probability
rates can be expressed in terms of permanents of positive
semi-definite matrices, depending on the interferometer evolution, the
spectral distribution of the sources and the times when the correlated
measurements occur. The permanent structure of these multi-order
probability rates is a manifestation of the underlying physics of multi-boson
interference and yields an interesting connection with the
so called boson sampling problem. 
\end{abstract}
\maketitle

\section{Motivation}

The Hanbury Brown and Twiss (HBT) experiment in $1956$ \cite{HanburyBrown1956}, aimed to measure the
angular size of a star by performing correlated detections, paved the way
towards the development of the field of quantum optics. From 1956 until now a
numerous series of remarkable experiments
\cite{Arecchi1966,Yasuda1996,Foelling2005,Schellekens2005,Agafonov2008,Liu2009,Hodgman2011,Oppel2012,Perrin2012} 
based on high order correlation measurements with thermal sources have been performed,
and important applications in high-precision imaging
\cite{Pittman1995,Bennink2002,Valencia2005,Ferri2005,Chan2009,Liu2009a,Zhou2010,Peng2014} and information processing
\cite{Tamma2014a} have been highlighted.

This fast advancement in experimental technologies based on thermal light
interferometry calls for a general description of multi-boson correlation
interferometry with thermal sources. Here we fully analyze HBT-like
experiments for arbitrary orders of correlation measurements, arbitrary passive
linear optical interferometers and arbitrary spectral distributions of the thermal sources.

Our analysis also brings up an interesting connection with the so-called \emph{Boson Sampling Problem} (BSP) 
\cite{aaronson2011computational,Franson2013,Ralph2013,Gard2014,Lund2014,Tamma2014},
where the probability of finding $\Ndets$ single input bosons in
$\Ndets\ll\Mports$ output ports of a $\Mports$-port interferometer depends on
permanents of random complex matrices
\cite{aaronson2011computational,Valiant1979}.

Differently from the BSP, multi-order correlation measurements at the output of
arbitrary interferometers rely additionally on the times the detections occur
\cite{tamma2014multi,Tamma2013,Naegele}. Further, for multi-mode thermal input sources, the
detection rates are connected with permanents of positive semi-definite matrices, whose
elements depend not only on the interferometer evolution but also on the average
rate of bosons emitted by each source and on the detection times.

Moreover, we show that these permanents arise from the interference of all multi-photon quantum paths
from the sources to the detectors. 

After giving a general perspective about \emph{Multi-Boson Correlation
Interferometry} (MBCI) with arbitrary sources in section~\ref{sec:MBCI}, we perform a full analysis for the case of thermal sources in section~\ref{sec:SingleFock}. 
In sections~\ref{sec:SingleMatrixFormulation} and \ref{sec:Equivalence}, we
derive two equivalent, interesting formulations of the
$\Ndets$-order correlation functions in terms of matrix permanents depending on
the interferometer evolution. Finally, we analyze the probability rates of
multi-order correlation measurements for approximatively equal detection times in section~\ref{sec:EqualDetectionTimes}, address the trivial case of thermal sources with equal average boson production rates in section~\ref{sec:Complexity} and conclude with final remarks in section~\ref{sec:Conclusion}. 

\section{Multi-Boson Correlation Interferometry (MBCI)}\label{sec:MBCI}
\begin{figure}
	\begin{center}
		\includegraphics{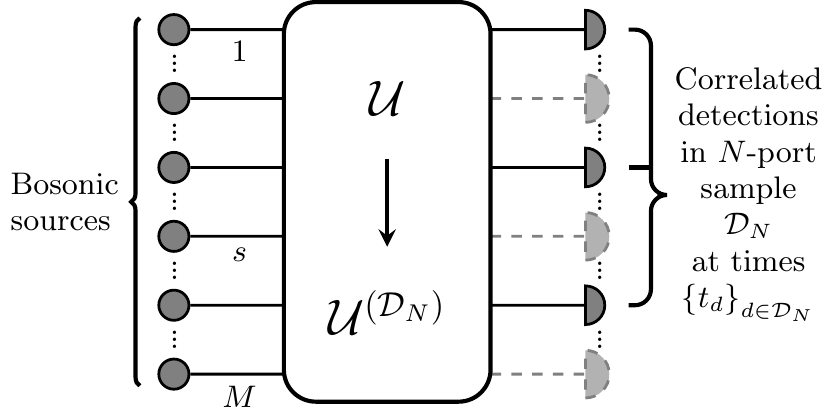}
		\caption{Multi-Boson Correlation Interferometry of order $N$ with a random
		linear interferometer with $M \geq N$ ports and bosonic sources. Here, we
		consider multi-mode thermal sources with arbitrary average boson rates
		$\meanrate{\sources} \ \sources=1,\dots,\Mports$.
	After the evolution in the interferometer, described by a unitary random
matrix $\UmatrixTot$, correlated detection events are recorded in the
$\Ndets$-port sample $\setD$ from the $M$ output ports independently of the remaining ports.}
		\label{fig:LinearInterferometerSingleFock}
	\end{center}
\end{figure}

The formulation of MBCI experiments of any given order $\Ndets$ is the following
(see Fig.~\ref{fig:LinearInterferometerSingleFock}): First, we prepare a 
linear $\Mports$-port interferometer with bosonic sources;
secondly, we consider
correlated detection events in which $\Ndets \leq \Mports$ single bosons are
detected in a $\Ndets$-port sample $\setD$ from the total $\Mports$ output
ports at joint detection times $\toutSet$, independently of the detection
outcomes for the remaining $\Mports - \Ndets$ detectors. 

We consider here the case of photonic sources, although our results can be easily extended to atomic interferometers with bosonic sources.
The probability rate for an $\Ndets$-fold joint detection event in a given
sample $\setD$ of output modes is proportional to the $\Ndets$th-order correlation function \cite{Glauber2007, Scully1997,Shih2011}
\begin{multline}
	\Gn{}  
	= \tr \left( \density{} \smashoperator[r]{\prod_{\detd\in\setD} }
	\Eoutminus{\detd} \smashoperator{\prod_{\detd\in\setD} }
	\Eoutplus{\detd}  \right),
	\label{eqn:GeneralSetup:CorrelationFunction}
\end{multline}
where $\Eoutplusminus{\detd}$ denotes the positive/negative frequency parts of the field operator $\Eout{\detd}=\Eoutplus{\detd} + \Eoutminus{\detd}$ at the $\detd$th detector. These field operators are connected with the field operators at the input ports 
by a unitary $\Mports \times \Mports$ matrix $\UmatrixTot$ describing the
interferometer, which we assume for simplicity to be frequency independent.
For a specific set $\setD$ of $\Ndets$ output ports where a joint detection occurs, the $\Ndets \times \Mports$ submatrix 
\begin{align}
	\UmatrixRect  &\defeq 
	\Big[ \Uto{\sources}{\detd} \Big]_{\begin{subarray}{l}\detd\in\setD \\
		\sources=1,\dots,\Mports \end{subarray}}
	\label{eqn:SingleFock:Umatrix}
\end{align}
of $\UmatrixTot$ allows us to express the electric field operators at the detectors as linear combinations 
\begin{align}
	\Eoutplus{\detd} = \sum_{\sources=1}^{\Mports} \Uto{\sources}{\detd}\Einplus{\sources}{\detd}
	\label{eqn:GeneralSetup:ExpansionEPlus}
\end{align}
of the field operators $\Einplus{\sources}{\detd}$ at the sources. 
Equivalent expressions hold for the conjugate fields $\Eoutminus{\detd}$. In the
next section we address MBCI experiments with multi-mode thermal states, while
we refer to \cite{tamma2014multi,TammaLaibacher1} for the case of multi-mode Fock states.

\section{MBCI with thermal input states}\label{sec:SingleFock}

One of the most natural optical sources in quantum optics is a thermal
source, which can be easily simulated in a laboratory by using, for example, a
laser beam impinging on a rotating ground glass \cite{Gonsiorowski1983}. Here, we consider the product state
\begin{align}
	\density{\text{th}} \defeq \bigotimes_{\sources=1}^{\Mports} \density{\sources}
	\label{eqn:TotalDensityMatrix}
\end{align}
of $\Mports$ independent multi-mode thermal states \cite{Glauber2007,Mandel1995}
\begin{align}
	\density{\sources} = \int \left[\prod_{\omega}^{} 
	\dtwo{\alpha_{\sources}(\omega)} \right] \Pfunct_{\sources,\text{th}}\left( \left\{ \alpha_{\sources}(\omega) \right\} \right) \bigotimes_{\omega} \ketbra{\alpha_{\sources}(\omega)}{\alpha_{\sources}(\omega)}
\end{align}
at each of the input ports $\sources=1,\dots,\Mports$, with Glauber-Sudarshan
$\Pfunct$-representation \cite{Sudarshan1963,Glauber1963}
\begin{align}
	\Pfunct_{\sources,\text{th}}(\left\{ \alpha_{\sources}(\omega) \right\}) \defeq \prod_{\omega}^{} \frac{1}{\pi \nmean{\sources}(\omega)} \exp \left( -\frac{\abs{\alpha_{\sources}(\omega)}^2}{\nmean{\sources}(\omega)} \right).
	\label{eqn:PRepresentation}
\end{align}
Here, the distribution $\nmean{\sources}(\omega) \defeq \meanrate{\sources}
\fdist_{\sources}(\omega)$ of the mean number of photons for the source
$\sources$ is defined by the normalized spectral distribution
$\fdist_{\sources}(\omega)$ and the mean rate $\meanrate{\sources}$ of photon
production. For simplicity, we assume equal Gaussian spectral distributions
\cite{Glauber2007}
\begin{align}
	\fdist(\omega) = \frac{1}{\sqrt{2\pi}\Delta\omega} \exp\left( -\frac{\left( \omega-\omega_0 \right)^2}{2\Delta\omega^2} \right)
\end{align}
with central frequency $\omega_0$ and bandwidth $\Delta\omega$, and their respective Fourier transform
\begin{align}
	\tdist(u) = \int_{-\infty}^{\infty}\d{\omega} \fdist(\omega) \ee{-\ii \omega u} = \ee{-\ii \omega_0 u} \exp\left(-\frac{u^2 \Delta\omega^2}{2}\right).
	\label{eqn:TDistGaussian}
\end{align}
For average photon rates $\meanrate{\sources}$ that are small compared to the
inverse of the time resolution of the detectors, the detection of more than one
photon in any of the output ports is very unlikely; thereby the use of photon
number resolving detectors is not necessary.

For the state~\eqref{eqn:TotalDensityMatrix}, Eq.~\eqref{eqn:GeneralSetup:CorrelationFunction} can be rewritten in terms of first order correlation functions
\begin{align}
	\Gonebare{}(\tout{\detd},\tout{\detd'}) \defeq
	\tr\left(\density{\text{th}}\,\Eoutminus{\detd}
	\Eoutplus{\detd'} \right)
	\label{eqn:FirstOrderCorrelationOutputDefinition}
\end{align}
as \cite{Glauber2007}
\begin{align}
	\Gn{}  
	= \sum_{\permut \in \SymmGroup{\Ndets}}^{} \prod_{\detd\in\setD} \Gonebare{} (\tout{\detd}, \tout{\permut(\detd)}),
	\label{eqn:GlauberFirstOrder}
\end{align}
where $\permut$ is an element of the symmetric group $\SymmGroup{\Ndets}$ of order $\Ndets$.

Since the different sources $\sources$ are independent, by defining
\cite{Glauber2007}
\begin{align}
	\Gonein{\sources}{\detd}{\detd'} &\defeq \Utoconj{\sources}{\detd} \Uto{\sources}{\detd'} 
	\tr\left( \density{\sources} \Einminus{\sources}{\detd}\Einplus{\sources}{\detd'} \right)
	 \nonumber \\
		&= \Econstant^2 \Utoconj{\sources}{\detd} \Uto{\sources}{\detd'} \meanrate{\sources} \thermaloverlap{\sources}{\detd}{\detd'},
		\label{eqn:FirstOrderCorrelation}
		\end{align}
where we used the narrow bandwidth approximation $\Delta\omega \ll \omega_0$
\footnote{In this case, the field operators can be approximated \cite{Loudon2000} as
	$\Einplus{\sources}{}(t) = \ii \Econstant \int_{-\infty}^{+\infty} \d{\omega}
	\ain{\sources}(\omega) \ee{-\ii \omega t}$
with the annihilation operators $\ain{\sources}(\omega)$ and a constant $\Econstant$.}, 
Eq.~\eqref{eqn:FirstOrderCorrelationOutputDefinition} becomes
\begin{align}
	\Gonebare{}(\tout{\detd},\tout{\detd'}) = \sum_{\sources=1}^{\Mports} \Gonein{\sources}{\detd}{\detd'}.
	\label{eqn:G1}
\end{align}
We point out that the $\Ndets$th order correlation function $\Gnbare{}$ in Eq.
(\ref{eqn:GlauberFirstOrder}) corresponds to the permanent of the matrix
$\big[\Gonebare{}(\tout{\detd},\tout{\detd'}) \big]_{\detd,\detd'}$ with
elements defined by Eqs.~\eqref{eqn:G1} and \eqref{eqn:FirstOrderCorrelation}. In the following sections, we derive two equivalent formulations of $\Gnbare{}$ in terms of matrix permanents depending on the entries of $\UmatrixRect$ in Eq.~\eqref{eqn:SingleFock:Umatrix}  and emphasize the underlying physics of multi-photon interference.

\subsection{\texorpdfstring{$\Ndets$}{N}th-order correlation functions and permanents}
\subsubsection{First formulation}\label{sec:SingleMatrixFormulation}
A compact expression of $\Gn{}$ in Eq.~\eqref{eqn:GlauberFirstOrder} can be obtained by defining the positive semi-definite matrix
\begin{align}
	\Cmatrix \defeq \Big[ \AmatrixEl{\detd}{\detd'}
	\thermaloverlap{}{\detd}{\detd'} \Big]_{\begin{subarray}{l}\detd\in\setD \\
		\detd'\in\setD \end{subarray}}.
\end{align}
Here $\AmatrixEl{\detd}{\detd'}$ are elements of the positive semi-definite matrix
\begin{align}
	\Amatrix \defeq \UmatrixRect  \operatorname{diag}\,(\meanrate{1},\dots,\meanrate{\Mports})\, \UmatrixRectAdj,
	\label{eqn:AmatrixDefinition}
\end{align}
while the positive semi-definite matrix
$\chi\defeq[\thermaloverlap{}{\detd}{\detd'}]_{\detd,\detd'\in\setD}$ describes
the pairwise degree of correlation of the $\Ndets$ detections depending on the detection times.

Moreover, the presence of both $\UmatrixRect$ and $\UmatrixRectAdj$ is  evidence of the multi-photon interference occurring in the optical network, as becomes clearer later.  
When we apply these definitions together with Eqs.~\eqref{eqn:FirstOrderCorrelation} and (\ref{eqn:G1}), Eq.~\eqref{eqn:GlauberFirstOrder}  becomes 
\begin{align}
	\Gn{}	
	= \Econstant^{2\Ndets} \per \Cmatrix.
	\label{eqn:GlauberFinal}
\end{align}
Thus, we find that the probability rate for an $\Ndets$-fold detection in a
given sample $\setD$ of output ports with thermal sources is mainly given by a single permanent of a positive semi-definite $\Ndets \times \Ndets$ matrix $\Cmatrix$.
From a physical point of view, while the matrix $\Amatrix$ contains the
interference-like terms associated with the interferometer evolution, the
time-dependent matrix $\chi$ accounts for the degree of correlation in time
between the different correlated measurements, as described in
Section~\ref{sec:EqualDetectionTimes}.

\subsubsection{Second formulation} \label{sec:Equivalence}
We just demonstrated that the correlation function $\Gnbare{}$ for a given
sample $\setD$ of output ports is proportional to the permanent of an $\Ndets
\times \Ndets$ matrix $\Cmatrix$. We notice that $\Cmatrix$ is not a  submatrix
of the unitary matrix $\UmatrixTot$ as in the case of the BSP with single photon
sources. We now show that $\Gnbare{}$ can also be expressed as a weighted sum of modulus squared permanents of matrices only built from columns of the interferometer submatrix $\UmatrixRect$ in Eq.~\eqref{eqn:SingleFock:Umatrix}. 

By substituting Eq.~\eqref{eqn:G1} in Eq.~\eqref{eqn:GlauberFirstOrder}, we obtain
\begin{align}
	\Gn{} &=  \sum_{\permut
		\in \SymmGroup{\Ndets}}^{} \prod_{\detd\in\setD}
		\sum_{\sources=1}^{\Mports} \Gonein{\sources}{\detd}{\permut(\detd)}.
	\label{eqn:GlauberToOwn1}
\end{align}
We now define the sets of ascending elements
\begin{align}
	\sourceSet
	= \left\{ \myunderbrace{1,\dots,1}{N_{1} \text{ times}}, \dots ,
	\myunderbrace{\sources,\dots,\sources}{N_{\sources} \text{ times}}, \dots ,
	\myunderbrace{\Mports,\dots,\Mports}{N_{\Mports} \text{ times}} \right\}
\vphantom{	\sourcevec{1}, \dots , \sourcevec{\Ndets} =
\myunderbrace{1,\dots,1}{N_{1} \text{ times}}, \dots ,
\underbrace{\sources,\dots,\sources}_{N_{\sources} \text{ times}}, \dots ,
\underbrace{\Mports,\dots,\Mports}_{N_{\Mports} \text{ times}}   },
	\label{eqn:GeneralCorrelations:ContrVector}
\end{align}
where $\Npart{\sources} \geq 0$ and $\sum_{\sources=1}^{\Mports} \Npart{\sources} = \Ndets$, with the associated weighting factors
\begin{align}
	\mathcal{N}(\sourceSet) \defeq \prod_{\sources=1}^{\Mports} \frac{1}{\Npart{\sources}!}.
\end{align}
These definitions allow us to write Eq.~\eqref{eqn:GlauberToOwn1} as
\begin{multline}
	\Gn{}	= \\
	\sum_{\sourceSet}^{} \mathcal{N}( \sourceSet ) \sum_{\permut \in
		\SymmGroup{\Ndets}} \sum_{\permutd \in \PathGroup}
		\prod_{\detd\in\setD} \Gonein{\permutd(\detd)}{\detd}{\permut(\detd)} ,
	\label{eqn:GnResummed}
\end{multline}
where $\PathGroup$ is the set of all $\Ndets!$ bijective functions that map the
set $\setD$ to the set $\sourceSet$. 
By using Eq.~\eqref{eqn:FirstOrderCorrelation} together with the matrices 
\begin{align}
	\ProdMatrix{\sourceSet}{\permut} \defeq \Big[ \Utoconj{c}{\detd}
	\Uto{c}{\permut(\detd)}\Big]_{\substack{\detd\in\setD \\
	c\in\sourceSet}},
	\label{eqn:ProdMatrixDefinition}
\end{align}
containing interference-like elements,
Eq.~\eqref{eqn:GnResummed} can be expressed as
\begin{widetext}
\begin{align}
	\Gn{} &= \Econstant^{2\Ndets} \sum_{\sourceSet}^{} \Bigg\{
		\mathcal{N}(\sourceSet) \left[
			\smashoperator[r]{\prod_{c\in\sourceSet}}
		\meanrate{c} \right]
		\sum_{\permut \in \SymmGroup{\Ndets}}^{} \left[
			\smashoperator[r]{\prod_{\detd\in\setD}} \thermaloverlap{}{\detd}{\permut(\detd)} \right] 
	\per \ProdMatrix{\sourceSet}{\permut} \Bigg\}.
	\label{eqn:OwnApproachExplicit}
\end{align}
\end{widetext}

The correlation function $\Gnbare{}$ in Eq.~\eqref{eqn:OwnApproachExplicit}
contains all contributions from the possible configurations $\sourceSet$  in
Eq.~\eqref{eqn:GeneralCorrelations:ContrVector} of ways the $\Ndets$ detected
photons can originate from the $\Mports$ sources. In particular, each
contribution has a weighting factor depending on the product of the respective
average photon rates $\meanrate{\sources}$.
Furthermore, each possible configuration $\sourceSet$ is associated with a
weighted sum  over $\permut$ (with weighting factors $\prod_{\detd\in\setD}
\tdist(\tout{\permut(\detd)} - \tout{\detd})$) of the permanents of the corresponding ``interference'' matrices  $\ProdMatrix{\sourceSet}{\permut}$. 

\subsection{Uncorrelated versus Correlated Detections}\label{sec:EqualDetectionTimes}

From the result in \eqref{eqn:GlauberFinal} it is evident that the pairwise degree of
correlation between the $\Ndets$ detections in an $\Ndets$-order correlation measurement 
is established by the positive semi-definite matrix
$\chi \defeq [\thermaloverlap{}{\detd}{\detd'}]_{\detd,\detd'\in\setD}$, whose
elements are defined by Eq.~\eqref{eqn:TDistGaussian}. Here, we will consider
the two extremal cases of completely uncorrelated or correlated detections.

In particular, the contribution to $\Gnbare{}$ in Eq.~\eqref{eqn:GlauberFinal}
by a given pair of detection events at detectors $\detd \neq \detd'$ vanishes if
$\abs{\tout{\detd} - \tout{\detd'}} \Delta\omega \gg 1$. if $\abs{\tout{\detd} - \tout{\detd'}} \Delta\omega \gg 1 \ \forall
\detd,\detd'$, which implies $\thermaloverlap{}{\detd}{\detd'} = \delta_{\detd,\detd'}$, and the only contributions to $\Gnbare{}$ are the ones for which
$\detd=\detd'$. In this case, Eq.~\eqref{eqn:GlauberFinal} trivially reduces to 
\begin{multline}
\Gnbare{}(\abs{\tout{\detd} - \tout{\detd'}} \Delta\omega \gg 1;\setD) = \\
 =\Econstant^{2\Ndets} \smashoperator[r]{\prod_{\detd\in\setD}}
\AmatrixEl{\detd}{\detd} 
= \smashoperator[r]{\prod_{\detd\in\setD}} \Gone{\detd}{\detd},
\end{multline}
where clearly the detections in the $\Ndets$ output ports are physically independent of each other and no multi-photon interference occurs.

On the other hand, in the condition of  approximately equal detection times
($\abs{\tout{\detd} - \tout{\detd'}} \Delta\omega \ll 1$), which implies $\prod_{\detd\in\setD} \tdist(\tout{\permut(\detd)} - \tout{\detd})= 1 \ \forall \permut \in \SymmGroup{\Ndets}$, Eq.~\eqref{eqn:GlauberFinal} simplifies to
\begin{align}
	\Gnbare{}(\abs{\tout{\detd} - \tout{\detd'}} \Delta\omega \ll 1;\setD) = \Econstant^{2\Ndets} \per \Amatrix,
	\label{eqn:GlauberFinalSameTimes}
\end{align}
which only depends on the mean photon rates of each source and on the interferometer transformation%
\footnote{After the completion of our work, the related independent research in
	\cite{Rahimi2014} came to our attention. Differently from the multi-mode
	thermal sources addressed in our paper, the authors consider monochromatic
	thermal sources, which correspond to the limit considered in
	Eq.~\eqref{eqn:GlauberFinalSameTimes}. Further they calculate the probability to find single
	photons in exactly $\Ndets$ of the $\Mports$ output ports and the vacuum
	in the others. Differently here we focus on the determination of
	experimental probability rates for correlated detections in an $\Ndets$-port sample $\setD$ at arbitrary time sequences $\toutSet$
	independently of the detection outcomes for the remaining ports. 
}.
Here, the complete interference between all possible $\Ndets$-photon multi-path contributions to a joint detection emerges from the permanent structure of the $\Ndets$th order correlation function. 

\begin{figure}
	\begin{center}

		\subfloat[$\mathcal{S}_2 = \left\{ a,b \right\}$]
		{\includegraphics[scale=0.99]{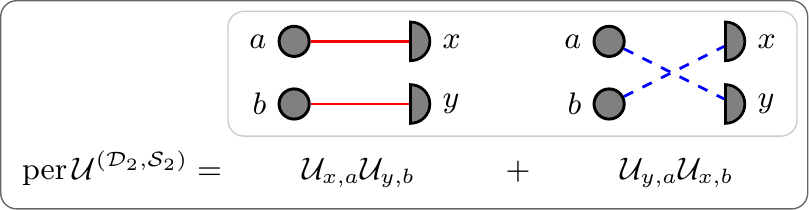}}

		\subfloat[$\mathcal{S}_2 = \left\{ a,a \right\}$]
		{\includegraphics[scale=0.99]{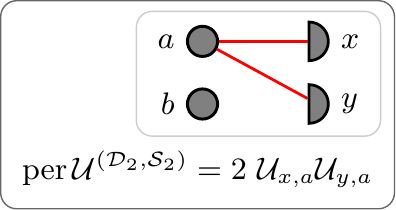}} \hspace{0.1cm}
		\subfloat[$\mathcal{S}_2 = \left\{ b,b \right\}$]
		{\includegraphics[scale=0.99]{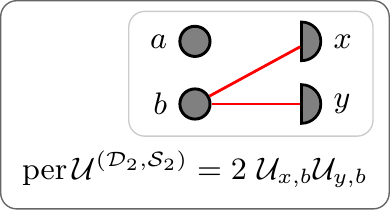}}
		\caption{Possible sets $\mathcal{S}_2$ of the source/s contributing to an
		$\Ndets$-fold detection at the $N=2$ ports of a given sample
		$\setD=\left\{ x,y \right\}$ from the $M$ interferometric output ports
		in Fig. 1, in the case of average photon rates $\meanrate{a}, \meanrate{b}
		\neq 0$ and $\meanrate{\sources} = 0 \ \forall \sources \neq a,b$. In set
	(a) both sources $a$ and $b$ contribute one photon, leading to two indistinguishable 2-photon quantum paths, each corresponding to a different term of the associated permanent. In sets (b) and (c), since both detected photons stem from a single source, only one 2-photon quantum path is possible, corresponding now to a single permanent term counted twice. Indeed, in both cases the associated matrix is constructed with two identical columns according to the contributing source.}
		\label{fig:QuantumPathExplanation}
	\end{center}
\end{figure}

In an analogous way, the equivalent expression of $\Gnbare{}$ in Eq.~\eqref{eqn:OwnApproachExplicit} simplifies to the incoherent sum
\begin{multline}
	\Gnbare{}(\abs{\tout{\detd} - \tout{\detd'}} \Delta\omega \ll 1;\setD) \approx \\
	\Econstant^{2\Ndets} 
	\sum_{\sourceSet}^{} \Bigg\{ \mathcal{N}(\sourceSet ) \left[
		\smashoperator[r]{\prod_{c\in\sourceSet}} \meanrate{c} \right] 
	\abs{\per \Umatrix{\sourceSet}}^2 \Bigg\} 
	\label{eqn:OursFinal}
\end{multline}
of weighted modulus squared permanents of the matrices
\begin{align}
	\Umatrix{\sourceSet} \defeq \Big[ \Uto{c}{\detd}
	\Big]_{\substack{\detd\in\setD \\ c\in\sourceSet}}.
	\label{MatricesGN}
\end{align}
Each matrix corresponds to a configuration $\sourceSet$ defining the number $\Npart{\sources}$ of photons each source contributes to the $\Ndets$-fold detection and can be obtained  by repeating each  column $s$ of the matrix $\UmatrixRect$  in Eq.~\eqref{eqn:SingleFock:Umatrix}  $N_s$ times.
The terms interfering in the modulus square of $\per \Umatrix{\sourceSet}$
correspond to all possible indistinguishable $N$-photon paths which connect the
$\Ndets$ sources $\sourceSet$ with the $\Ndets$ detectors of a given sample
$\setD$, as illustrated in Fig.~\ref{fig:QuantumPathExplanation} in the case $N=2$.

 In general, the lower the column repetition rate in Eq. (\ref{MatricesGN}) is for a given configuration $\sourceSet$, the higher is the number of physically interfering $\Ndets$-photon quantum paths and the corresponding \textit{degree of multi-photon interference}.
In particular, the only configurations where no column repetition occurs are the ones where $\Ndets$ sources contribute to an $\Ndets$-fold detection (see Fig. 2 (a) for $N=2$), as in the original boson sampling formulation with single-photon sources. Indeed, these configurations correspond to $N!$ interfering $N$-photon paths.

\subsection{Equal average photon rates} \label{sec:Complexity}
 We now consider the trivial case where all thermal sources have mean photon
 rates $\meanrate{\sources} = \meanrate{} \ \forall \sources$ and derive two notable properties for the permanents of the matrices $\ProdMatrix{\sourceSet}{\permut}$  in Eq. (\ref{eqn:ProdMatrixDefinition}) and $\Umatrix{\sourceSet}$ in Eq. (\ref{MatricesGN}). 
 
 In this case, we easily find that the correlation function in Eq. (\ref{eqn:GlauberFinal}) reduces to the constant expression
\begin{align}
	\Gn{} = \Econstant^{2\Ndets} \meanrate{}^{\Ndets} 
	\label{eqn:CorrelationEqualMeann},
\end{align}
which, as expected \cite{Weedbrook2012}, is independent of the evolution in the interferometer. 
If we compare Eq.~\eqref{eqn:CorrelationEqualMeann} with
Eq.~\eqref{eqn:OwnApproachExplicit} in the limit of identical mean photon rates, we find that the property
\begin{align}
	\sum_{\sourceSet}^{} \mathcal{N}(\sourceSet) \per \ProdMatrix{\sourceSet}{\permut} = 
	\begin{dcases}
		1 & \permut = \unity \\
		0 & \permut \neq \unity
	\end{dcases}
\end{align}
holds for the matrices $\ProdMatrix{\sourceSet}{\permut}$ in Eq.
(\ref{eqn:ProdMatrixDefinition}). Further, since
Eq.~\eqref{eqn:CorrelationEqualMeann} is independent of the detection times
$\tout{\detd}$, it must also correspond to the expression~\eqref{eqn:OursFinal}
in the condition of equal mean photon rates. This yields the second property
\begin{align}
	\sum_{\sourceSet}^{} \mathcal{N}(\sourceSet) \abs{\per \Umatrix{\sourceSet}}^2 = 1
\end{align}
for the matrices $\Umatrix{\sourceSet}$ in Eq. \eqref{MatricesGN}. These two
properties arise since the photon-counting probability rates for sources with
equal average intensity are physically independent from the interferometer.

\section{Final remarks}\label{sec:Conclusion}

We performed a full analysis of multi-boson correlation interferometry of arbitrary order $N\leq M$, where $M$ are the ports of a random passive linear interferometer, for thermal sources with arbitrary spectral distributions.

We showed that the probability rates of detecting single bosons
in at least $N$ output ports, with $N\leq M$,  are  proportional to the
permanents of positive semi-definite $\Ndets \times \Ndets$ matrices, leading to an
interesting connection with the boson sampling problem. Each matrix is given by
the Hadamard product (product of the corresponding entries) of a
time-dependent matrix, describing the degree of correlation in time between the
measurements, and the interference-dependent matrix associated with the
interferometer evolution and the average photon rate of each source. 

Moreover, we demonstrated that, for approximately equal detection times, the
$\Ndets$-boson probability rates can be cast as a time-dependent weighted sum of modulus squared permanents of matrices with interference-like elements depending only on the interferometer evolution. 
Indeed, each different permanent is associated  with a possible physical
configuration for the number of bosons each source contributes to the detection
and describes the interference of all the corresponding multi-boson quantum paths 
from the sources to the detectors. The higher the number of sources contributing
to the joint detection is, the  larger the number of corresponding interfering
multi-path amplitudes is.

In conclusion, our general analysis of multi-boson correlation interferometry
with thermal sources provides a deeper insight in the fundamental physics of
multi-boson interference for arbitrary order HBT-like experiments where highly
interesting correlation effects emerge.

\begin{acknowledgments}
V.T. would like to thank  M. Freyberger, F. N\"{a}gele, W. P.
Schleich, and K. Vogel, as well as J. Franson, S. Lomonaco, T. Pittmann,  and
Y.H. Shih for fruitful discussions during his visit at UMBC in the
summer of 2013.

V.T. acknowledges the support of the German Space Agency DLR with funds
provided by the Federal Ministry of Economics and Technology (BMWi) under
grant no. DLR 50 WM 1136.

This work was also supported by a grant from the Ministry of Science, Research and the Arts of Baden-W\"urttemberg (Az: 33-7533-30-10/19/2).
\end{acknowledgments}
\end{document}